Roman Kusche*, Jan Graßhoff, Andra Oltmann, Lukas Boudnik and Philipp Rostalski

# A Robust Multi-Channel EMG System for Lower Back and Abdominal Muscles Training

**Abstract:** EMG is an established method to acquire the action potentials of contracted muscles. Although commercial EMG systems are available and it is one of the most researched biosignals, it has never become widely used in rehabilitation or fitness training monitoring. The reasons are technical challenges of wearable EMG systems regarding electrode placement, motion artefacts and the complex connectivity of multi-channel EMG measurements. We address this problem for the lower back and abdominal musculature, through a novel dry electrodes belt, multi-channel high density EMG circuitry and problem-specific signal processing. The subject can easily strap the dry electrodes belt around himself which provides 16 EMG channels. Interferences from the ECG and motion artefacts are reduced by a stationary wavelet decomposition. Afterwards, an inter-channel filter is applied to increase the robustness of the signals. Subject measurements during different kinds of typical abdominal and lower back training exercises were performed wearing the novel dry electrodes belt. The results show the possibility of robust EMG measurements from the lower back and abdominal muscles by utilizing the gathered redundancy, appropriately. The additional information obtained via the multi-channel EMG circuitry and spatial oversampling can be used to address current problems of EMG applications. It combines the advantages of robustness and the capability of using comfortable dry electrodes. Therefore, the proposed measurement method for acquiring spatial information about the muscle contractions from the lower trunk can be used for rehabilitation or fitness training monitoring.

**Keywords:** Dry electrodes, electrodes belt, EMG array, surface EMG, training monitoring, wearable.

____________

**\*Corresponding author: Roman Kusche:** Fraunhofer Research Institution for Individualized and Cell-Based Medical Engineering IMTE, Mönkhofer Weg 239a, Lübeck, Germany, e-mail: roman.kusche@imte.fraunhofer.de
**Jan Graßhoff, Andra Oltmann, Lukas Boudnik, Philipp Rostalski:** Fraunhofer Research Institution for Individualized and Cell-Based Medical Engineering, Mönkhofer Weg 239a, Lübeck, Germany

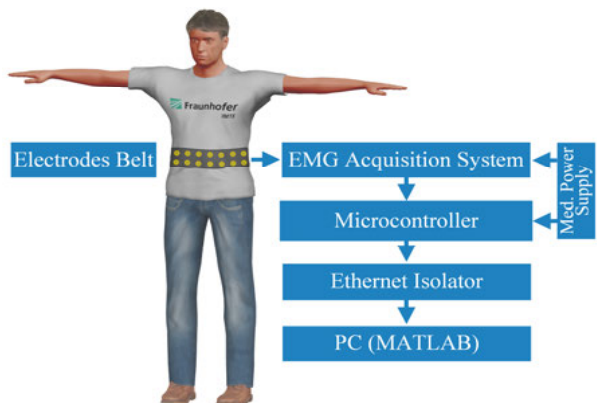

**Figure 1:** Principle of the multi-channel EMG data acquisition from the lower back and abdominal muscles.

## 1 Introduction

Lower back and abdominal muscles are responsible for stabilizing and moving the upper part of the human body [1]. A balanced training is therefore aimed at both sports and rehabilitation. To get not only a visual impression about the training performance, spatial detection of muscle contractions can be useful [2]. For that, electromyography (EMG) is an established measurement technique. However, the placement of many adhesive gel electrodes, significant motion artefacts and strong influences from the superimposed ECG signal are challenging [3, 4]. To address these problems of multi-channel EMG measurements, we propose a novel system, based on the combination of a comfortable dry electrodes belt, a high-resolution measurement circuitry and an adaptive signal processing chain. The system's capability to acquire spatially resolved lower back and abdominal muscle contractions is demonstrated via human subject measurements.

## 2 Materials and Methods

### 2.1 Signal Conditions

Under relaxed laboratory conditions, two electrical voltage signal components, ECG and EMG, are dominant in the torso.

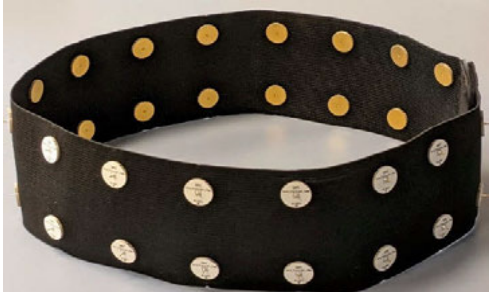

**Figure 2:** Stretchable dry electrodes belt for comfortable and equidistant electrode placement.

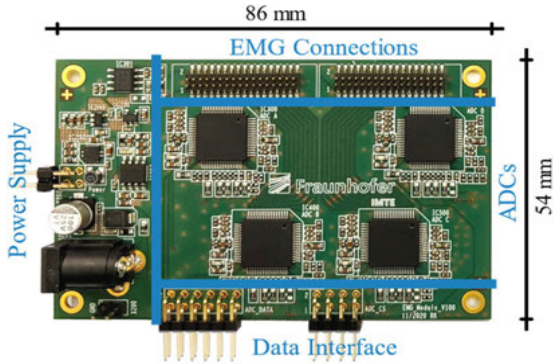

**Figure 3:** Printed circuit board of the EMG acquisition system, capable of digitizing 16 channels, simultaneously.

Their amplitudes depend on the specific electrode positions, but are both in the range of several hundreds of µV [5]. A wide overlap of the frequency ranges of both signals makes separations challenging [5]. In addition, the wanted EMG signal is typically distorted by noise from the mains and unpredictable motion artefacts, especially during sports or rehabilitation exercises.

## 2.2 Data Acquisition

In order to address the described challenges, a novel problem-specific signal acquisition setup has been developed, as illustrated in Fig. 1. It consists of an electrodes belt, an electrical EMG acquisition system and a microcontroller system for interfacing and controlling purposes. To protect the patient from the mains supply, these components are powered by a medical power supply (MPU31-102, SINPRO Electronics, Pingtung City, TW) and the ethernet connection to a PC is galvanically isolated (MED MI 1005, Baaske Medical, Lübbecke, DE).

Since the usage of commonly used adhesive gel electrodes is uncomfortable and may cause skin irritations, 32 gold-plated dry electrodes with a diameter of 15 mm are placed on a stretchable belt, as shown in Fig. 2 [3, 6]. Using a flat band cable, the electrical signals of the resulting 16 differential vertical EMG signals are connected to the EMG acquisition system.

The main functions of the acquisition system are amplification and high-resolution digitization of the measurement channels, achieved via synchronized 24-bit analogue-to-digital converters (ADC, ADS131E08, Texas Instruments, TX, USA) with a sampling rate of $f_s$=4 kSPS. To reduce the 50/60 Hz interferences from the mains, an additional common mode rejection circuit is implemented. A photograph of the developed 4-layers printed circuit board (PCB), which has credit card dimensions, is shown in Fig. 3.

The digitised data is transmitted via a Serial Peripheral Interface (SPI) to an external microcontroller board. The 32-bit 600 MHz ARM Cortex-M7 controller pre-processes the data and forwards it via a native 100 Mbit/s ethernet interface to a PC for signal processing and analysis.

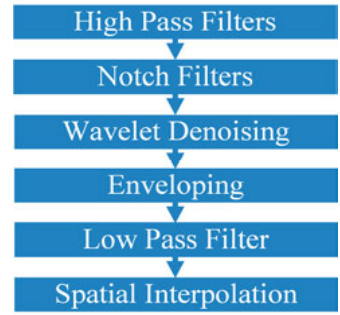

**Figure 4:** Principle of the digital signal processing.

## 2.3 Signal Processing

The signal processing chain, which can be seen in Fig. 4, is implemented in MATLAB from The MathWorks and performed on all 16 EMG channels.

First, low frequency components, especially caused by drifting effects, are removed via a high pass filter (N=1; $f_c$=10 Hz). Before continuing with the next steps, the signal is checked for reasonable amplitude values to detect potential bad electrode contact conditions. If the amplitudes are implausibly high, it is interpreted as a disturbance and the signal is discarded 0. In the second step, the interferences caused by the mains are removed using notch filters with notch frequencies of $f_N$={50; 100; 150} Hz. Cardiac artefacts are then removed using a wavelet denoising method similar to [7]. The signal is decomposed using the Stationary Wavelet

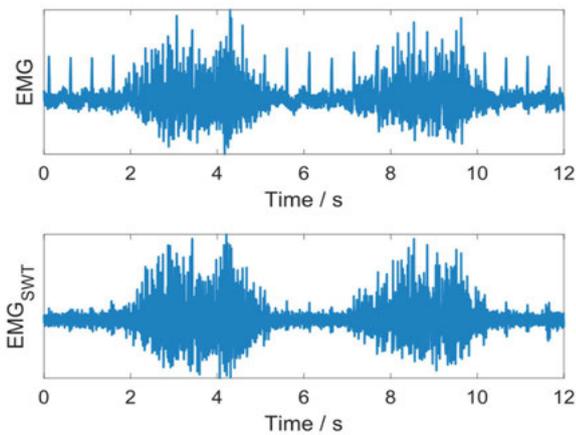

**Figure 5:** Exemplary EMG signal before and after ECG removal via wavelet denoising.

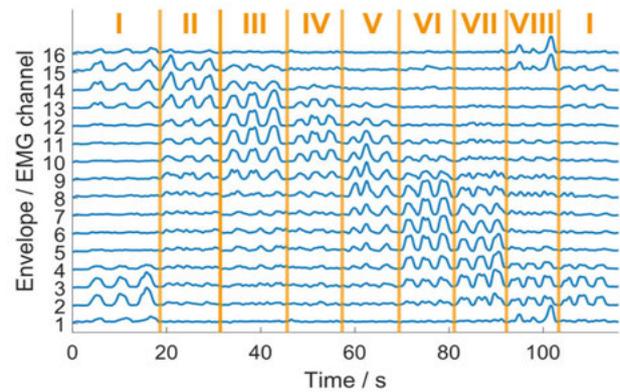

**Figure 6:** Envelopes of the 16 EMG channels during the stepwise contraction of different muscle regions.

Transform (SWT) implemented as a 7-level filter bank with 4th order Daubechies wavelets and ECG artefacts are removed via level-dependent hard-thresholding. Afterwards, the envelopes of the extracted EMG signal components are calculated and low pass filtered (N=2, $f_c$=0.5 Hz). Since motion artefacts are difficult to separate from actual muscle contractions, the robustness of the interpretation is increased by utilizing the high redundancy of the multi-channel approach. This is technically realized by averaging of each EMG channel with the respective two adjacent channels.

## 2.4 Measurement Protocol

To demonstrate the performance of this system, three subject measurements are performed. Prior to each measurement, the gold-plated dry electrodes are cleaned with ethanol. However, the skin is not prepared in any way. The stretchable electrodes belt is then placed around the upper body at the height of the umbilicus in a comfortable way. Since the contact conditions are poor when using dry electrodes, the belt is worn for 5 minutes before beginning with the actual measurements to enable the accumulation of fluids under the electrodes, working as electrolytes.

## 3 Results

For the first measurement, the subject is asked to stand straight and to tilt the upper body 3 times in one direction by approximately 20°-30°, starting with bending forward (I). Afterwards, the subject repeats this exercise in the front-left (II), left (III), back-left (IV), back (V), back-right (VI), right (VII), front-right (VIII) and finally again the front (I) direction. To demonstrate the effect of the adaptive wavelet denoising, one exemplary resulting EMG signal is shown in the upper plot of Fig. 5. In this representation, the typical R-peaks are clearly visible. Below, the same signal interval after the denoising is shown. It can be seen, that the ECG signal components are significantly attenuated. In Fig. 6, the envelopes of all 16 EMG measurement channels are shown for the entire measurement procedure after the complete signal processing chain. In all plots, the intervals of three repeated exercises can be recognised. As expected, the focus of EMG activity also shifts with each interval from EMG channel 16 (back) left around the body to channel 1 (back).

For the second measurement, the subject is asked to lie on the back and do sit-ups, as illustrated in the upper part of Fig. 7. Instead of focussing on the time series, the subjacent stem plot shows the EMG signal intensities of all 16 channels at the point in time when the muscle contractions are at their maxima. A clear directionality towards the frontcan be observed, which decreases towards the back. Although strong mechanical influences on the electrodes on the back are to be expected during this exercise, they are not visible in the result due to the filtering and interpolation presented before. For better visualization, the bottom figure graphically projects these results on an upper body model.

During the last measurement, the subject lies on his stomach and contracts the back muscles, as illustrated in the upper part of Fig. 8. Complementary to the previous measurement, a directionality towards the back can be observed, which decreases towards the front. This expected behaviour is also visible in the corresponding graphical representation below. However, it is apparent that two channels at the back only have very weak signal components. This is caused by electrode lift-ups and the described fault detection.

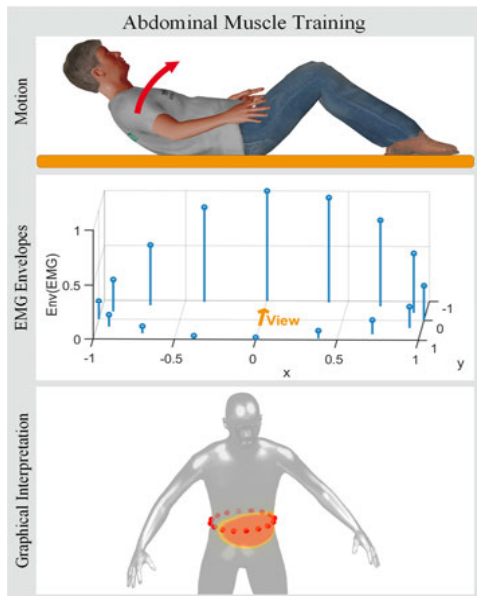

**Figure 7:** Measurement results of the abdominal muscle training.

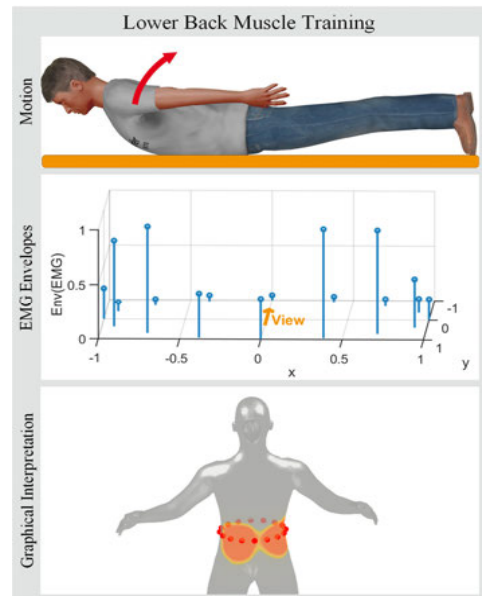

**Figure 8:** Measurement results of the lower back muscle training.

## 4 Discussion

The performed exercises are realistic motion sequences in which strong artefacts affect the electrodes. It is shown that these do not have a significant influence on the measurement results and that the spatial distribution of the EMG activities is well identifiable. However, more extensive studies will be needed in the future for further analysis and discussion.

## 5 Conclusion

The combination of problem-specific measurement technology, spatial oversampling and adaptive signal processing enables robust EMG measurements even during fitness or rehabilitation exercises. In the future, further subject studies will be performed to identify new possibilities and limitations of this approach.

**Author Statement**
Research funding: This work was supported by European Union – European Regional Development Fund, the Federal Government and Land Schleswig-Holstein, Project: "Diagnose- und Therapieverfahren für die Individualisierte Medizintechnik (IMTE)", Project No. 12420002. Conflict of interest: Authors state no conflict of interest. Informed consent: Informed consent has been obtained from all individuals included in this study. Ethical approval: The research related to human use complies with all the relevant national regulations, institutional policies and was performed in accordance with the tenets of the Helsinki Declaration, and has been approved by the authors' institutional review board (declaration number 21-020).